\documentstyle[amssymb,12pt,epsf,epsfig]{article}

\begin{document}

\title{Quark Correlations in Nucleons and Nuclei}
\author{G. Musulmanbekov }
\date{}
\maketitle
\begin {abstract} A dynamical quark model of hadron and nucleus structure is
proposed. In the frame of the model, called the Strongly Correlated Quark
Model, quarks and nucleons inside nuclei are arranged in crystal--like
structure.
\end {abstract}

\bigskip
How are nucleon properties modified inside nuclei and do quarks manifest
themselves explicitly in ground state nuclei? These questions are the
topics of this paper, where we show that the nuclear modification of
nucleon properties is a result of correlations of quarks of bound
nucleons.

The analysis has been performed in the frame of the so--called Strongly
Correlated Quark Model (SCQM) \cite{Mus1}. \ The ingredients of the model
are the following. A single quark of definite color embedded in the vacuum
begins to polarize its surrounding resulting in the formation of quark and
gluon condensate. At the same time it experiences the pressure of the
vacuum because of zero point radiation field or vacuum fluctuations which
act on the quark tending to destroy the ordering of the condensate.
Suppose that we place the corresponding antiquark in the vicinity of the
first one. Owing to their opposite signs color polarization fields of the
quark and antiquark interfere destructively in the overlapped space
regions eliminating each other maximally in the space around the
middle--point between the quarks. This effect leads to a decrease in
condensates density in that space region and overbalancing of the vacuum
pressure acting on the quark and antiquark from outer space regions. As a
result an attractive force between the quark and antiquark emerges and the
quark and antiquark start to move towards each other. The density of the
remaining condensate around the quark (antiquark) is
identified with the hadronic matter distribution. At maximum displacement in the $%
\overline{q}q-$ system corresponding to small overlapping of polarization
fields, hadronic matter distributions have maximum extent and densities.
The closer they come to each other, the larger is the destructive
interference effect and the smaller hadronic matter distributions around
quarks and the larger their kinetic energies. In that way quark and
antiquark start to oscillate around their middle--point. For such
interacting $\overline{q}q-$ pair located on the $X$ axis at a distance
$2x$ from each other, the total Hamiltonian is

\begin{equation}
H=\frac{m_{\overline{q}}}{(1-\beta ^{2})^{1/2}}+\frac{m_{q}}{(1-\beta
^{2})^{1/2}}+V_{\overline{q}q}(2x),
\end{equation}
were $m_{\overline{q}}$, $m_{q}$ are the current masses of the valence
antiquark and
quark, $\beta =\beta (x)$ is their velocity depending on displacement $x$, and $%
V_{\overline{q}q}$ is the quark--antiquark potential energy with
separation $2x.$ It can be rewritten as
\begin{equation}
H=\left[ \frac{m_{\overline{q}}}{(1-\beta ^{2})^{1/2}}+U(x)\right] +\left[
\frac{m_{q}}{(1-\beta ^{2})^{1/2}}+U(x)\right] =H_{\overline{q}}+H_{q},
\end{equation}
were $U(x)=\frac{1}{2}V_{\overline{q}q}(2x)$ is the potential energy of
quark or antiquark. The quark (antiquark) with the surrounding cloud
(condensate) of quark -- antiquark pairs and gluons, or hadronic matter
distribution, forms the constituent quark. It is natural to assume that
the potential energy of the quark (antiquark), $U(x),$ corresponds to the
mass $M_{Q}$ of the constituent quark:
\begin{equation}
2U(x)=C_{1}\int_{-\infty }^{\infty }dz^{\prime }\int_{-\infty }^{\infty
}dy^{\prime }\int_{-\infty }^{\infty }dx^{\prime }\rho (x,{{\bf r}^{\prime }}%
)\approx 2M_{Q}(x)
\end{equation}
where $C_{1}$ is a dimensional constant and the hadronic matter density
distribution, $\rho (x,{{\bf r}^{\prime }}),$ is defined as
\begin{equation}
\rho (x,{{\bf r}^{\prime }})=C_{2}\left| \varphi (x,{\bf r}^{\prime
})\right| =C_{2}\left| \varphi _{Q}(x^{\prime }+x,y^{\prime },z^{\prime
})-\varphi _{\overline{Q}}(x^{\prime }-x,y^{\prime },z^{\prime })\right| .
\end{equation}
Here $C_{2}$ is a normalization constant, $\varphi _{Q}$ and $\varphi
_{\overline{Q}}$ are density profiles of the condensates around the quark
and antiquark located at distance $2x$ from each other. We consider by
convention the condensates around quark and antiquark having opposite
color charges. They have properties similar to compressive stress and
tensile stress (around defects) in solids.
Generalization to three--quark system in baryons is performed according to $%
SU(3)_{color}$ symmetry: in general, pair of quarks have coupled
representations
\begin{equation}
3\otimes 3=6\oplus \overline{3}
\end{equation}
in $SU(3)_{color}$ and for quarks within the same baryon only the $\overline{%
3}$ (antisymmetric) representation occurs. Hence, an antiquark can be
replaced by two correspondingly colored quarks to get a color singlet
baryon and destructive interference takes place between color fields of
three valence quarks (VQs). Putting aside the mass and charge differences
of valence quarks we may say that inside the baryon three quarks oscillate
along the bisectors of equilateral triangle. Therefore, keeping in mind
that the quark and antiquark in mesons and the three quarks in baryons are
strongly correlated, we can consider each of them separately as undergoing
oscillatory motion under the potential (3) in 1+1 dimension. Hereinafter
we consider VQ oscillating along the $X-$ axis, with $Z-$ axis
perpendicular to the plane of oscillation $XY$. Density profiles of
condensates around VQs are taken in gaussian form. This choice is dictated
by our semiclassical treatment of VQs motion. It has previously been shown
\cite{sch} that the wave packet solutions of the time dependent
Schrodinger equation for the harmonic oscillator move in exactly the same
way as corresponding classical oscillators. These solutions are called
''coherent states''. This relationship justifies (partly) our
semiclassical treatment of quantum objects.

We define the mass of constituent quark at maximum displacement as
\[
M_{Q(\overline{Q})}(x_{\max })=\frac{1}{3}\left( \frac{m_{\Delta }+m_{N}}{2}%
\right) \approx 360\ MeV,
\]
where $m_{\Delta }$ and $m_{N}$ are masses delta--isobar and nucleon
correspondingly. The parameters of the model, namely, maximum displacement, $%
x_{\max },$ and parameters of the gaussian function, $\sigma _{x,y,z},$
for hadronic matter distribution around VQ are chosen to be
\begin{equation}
x_{\max }=0.64\ fm,\ \sigma _{x,y}=0.24\ fm,\ \sigma _{z}=0.12\ fm.
\end{equation}
They are adjusted by comparison of calculated and experimental values of
inelastic cross sections, $\sigma _{in}(s),$ and inelastic overlap function $%
G_{in}(s,b)$ for $pp$ and $\overline{p}p-$ collisions \cite{Mus2}. The
current mass of the valence quark is taken to be $5\ MeV$. The behavior of
potential (3) demonstrates the relationship between constituent and
current quark states inside a hadron (Fig. 1). At maximum displacement
quark is a nonrelativistic, constituent one (VQ surrounded by condensate),
since the influence of polarization fields of other quarks becomes minimal
and the VQ possesses the maximal potential energy corresponding to the
mass of the constituent quark. At the origin of oscillation, $x=0,$ the
antiquark and quark in mesons and the three quarks in baryons, being close
to each other, have maximum kinetic energy and correspondingly minimum
potential energy and mass: they are relativistic, current quarks (bare
VQs). This configuration corresponds to so--called ''asymptotic freedom''.
In the intermediate region there is increasing (decreasing) of the
constituent quark mass by dressing (undressing) of VQs due to decreasing
(increasing) of the destructive interference effect. This mechanism meets
the local gauge invariance principle. Indeed, destructive interference of
color fields of the quark and antiquark in mesons and three quarks in
baryons depending on their displacements can be treated as phase rotation
of wave function of single VQ in color space $\psi _{c}$ on angle $\theta
$ depending on displacement $x$ of the VQ in coordinate space
\begin{equation}
\psi _{c}(x)\rightarrow e^{ig\theta (x)}\psi _{c}(x).
\end{equation}
Phase rotation,in turn, leads to VQ dressing (undressing) by the quark and
gluon condensate that corresponds to the transformation of the gauge field
\begin{equation}
A_{\mu }(x)\rightarrow A_{\mu }(x)+\partial _{\mu }\theta (x).
\end{equation}
Here we drop the color indices of $A_{\mu }(x)$ and consider each quark of
specific color separately as changing its effective color charge, $g\theta
(x),$ in color fields of other quarks (antiquark) due to destructive
interference. Thus gauge transformations (7, 8) map internal (isotopic)
space of the colored quark onto coordinate space. On the other hand this
dynamical picture of VQ dressing (undressing) corresponds to chiral
symmetry breaking (restoration). Due to this mechanism of VQs
oscillations, the nucleon runs over the states corresponding to the
certain terms of the infinite series of Fock space
\begin{equation}
\mid B\rangle=c_{1}\mid q_{1}q_{2}q_{3}\rangle+c_{2}\mid q_{1}q_{2}%
q_{3}\overline{q}q\rangle+c_{3}\mid q_{1}q_{2}q_{3}g\rangle...
\end{equation}
The proposed model has some important consequences. Inside hadrons quarks
and their accompanying gluons, as well, are strongly correlated. Nucleons
are nonspherical objects: they are flattened along the axis perpendicular
to the plane of quarks oscillations. First, because VQs undergo plane
oscillations and second, owing to the flatness of hadronic matter
distributions around VQs (according to (6)).
\begin{figure}[tb]
\begin{center}
\epsfig{figure=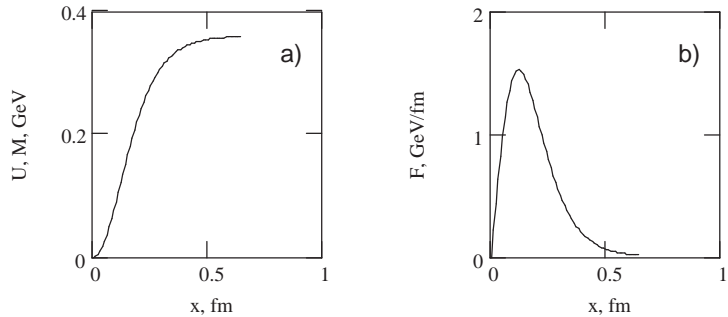,width=4.5in}
\end{center}
\caption{ a) Potential energy of valence quark and mass of constituent
quark; b) ''Confinement'' force.}
\end{figure}

From the form of the quark potential (Fig.1) one can conclude that the
dynamics of VQ corresponds to a nonlinear oscillator and VQ with its
surrounding can be treated as a nonlinear wave packet. Moreover, our quark
-- antiquark system turns out to be identical to the so--called
''breather'' solution of the sine--Gordon (SG) equation\cite{Mus3}. SG
equation in (1+1) dimension in reduced form for scalar function $\phi
(x,t)$ is given by
\begin{equation}
\Box \phi (x,t)+\sin \phi (x,t)=0,
\end{equation}
where $x,$ $t$ are dimensionless. The breather is a periodic solution
representing bound state of soliton--antisoliton pair which oscillates
around their center of mass:
\begin{equation}
\phi _{br}(x,t)=4\tan ^{-1}\left[ \frac{\sinh \left[ ut/\sqrt{(1-u^{2}}%
\right] }{u\cosh \left[ x/\sqrt{(1-u^{2}}\right] }\right] ,
\end{equation}
where $u$ is 4--velocity. During the oscillations of the
soliton--antisoliton pair their density profile
\begin{equation}
\varphi _{s-as}(x,t)=\frac{d\phi _{br}(x,t)}{dx}
\end{equation}
evolves like our quark--antiquark system, i.e. at maximal displacement the
soliton and antisoliton are maximal and at minimum displacement they
''annihilate''. This similarity is not surprising because our
quark--antiquark system was formulated in close analogy with the model of
dislocation--antidislocation \cite{frenkel}, which in its continuous limit
is described by the breather solution of the SG equation. It can be shown
that the soliton, antisoliton and breather obey relativistic kinematics,
i.e. their energies, momenta and shapes are transformed according to
Lorentz transformations. Thus the dynamics of our strongly correlated
quark system can be described by a sine--Gordon equation. Since the above
consideration of quarks as solitons is purely classical, an important
problem is to construct \ quantum states around them. Although the soliton
solution of the SG equation looks like an extended (quantum) particle, the
relation between classical solitons and quantum particles is not so
trivial. Techniques for quantization of classical solitons with usage of
various methods has been developed by many authors. The most well known of
them is semiclassical method of quantization (WKB) which allows one to
relate classical periodic orbits (breather solution of SG) with the
quantum energy levels \cite{DHN}.

So far we have dealt with the scalar polarization field around VQ. How can
one include spin in the frame of these classical considerations? According
to the prevailing belief, the spin is a quantum feature of microparticles
and has no classical analog. However, Belinfante \cite{belin} as early as
in 1939 showed that the spin of an electron may be regarded as an angular
momentum generated by a circulating flow of energy, or a momentum density,
in the wave field of the electron. Furthermore, a comparison between
calculations of angular momentum in the Dirac and electromagnetic fields,
performed by Ohanian \cite{ohanian}, shows that the spin of the electron
is entirely analogous to the angular momentum carried by classical
circularly polarized waves. Inclusion of spin in our scheme brings us to
spinning quarks--solitons or extended vortex representation of constituent
quarks. As shown in a previous paper \cite {Mus3} the dominating
contribution to proton spin comes from the orbital angular momentum of
gluons and sea $\overline{q}q-$pairs circulating around the oscillating
VQs. Gauge field $A_{\mu }(x)$ in (8) contains along with the scalar part,
$\varphi $, vector components, ${\bf A}$, as well.

Now let us proceed to the many--nucleon problem. What would happen with
oscillating\ quarks if we place a proton and a neutron nearby? Suppose
that they aligned occasionally in such a way that a pair of quarks (one
quark from the proton and another one from the neutron) with different
flavor and color are nearest. Vacuum pressure acting from outside on these
quarks decreases because of mutual influence (destructive interference,
again) of the different color fields of these quarks. This effect results
in attractive force between the nearest quarks from the proton and
neutron. This force is near half in magnitude for that between the quark
and antiquark in mesons and the three quarks in baryons at the same
displacements. An additional attractive force comes from the flavor
difference. To restrict these attractive forces quarks need to have
parallel spins. As a result the potential (3) for adjacent quarks acquires
an additional minimum at large quark displacements, small compared with
the primary one. In that way the proton and nucleon form a bound state,
namely, a deuteron. Now quarks of both nucleons oscillate around the
deuteron center of mass interchanging their positions, with the quarks at
free ends possessing maximal displacements. When all quarks pass the
deuteron center of mass the adjacent quarks (at proton -- neutron linkage)
acquire angular orbital momentum ($l=2$) that should be accompanied with
spin flip of both quarks to conserve the total angular momentum of the
deuteron.

Noting that three quarks inside nucleons are totally antisymmetric in the
color space and two quarks from different nucleons at linkage are in
antisymmetric color state ( $\overline{3}$) having different flavors and
parallel spins, we can construct more complex nuclei. The three nucleon
system is formed by linkage of two quarks of each nucleon with quarks of
two other nucleons according to the above rules. Three nucleon nuclei,
namely $^{3}H$ and $^{3}He$, represent a triangular configuration with
three quarks at free ends. There is a drastic difference of quark
oscillation pattern starting from the three nucleon system. The decrease
of external vacuum pressure on quarks in each nucleon results in decreased
attraction force between them that leads to the displacement of the
origins of oscillations of each quark to the nucleon periphery and to the
amplitude reduction of quarks oscillations, i.e. each quark having
constituent mass oscillate near its own origin of oscillation. And so the
suppression of current (bare) quark configurations inside the nuclear
medium occurs. Completion of a four--nucleon system, $^{4}He$, from a
three--nucleon one, occurs by binding free quark ends in $^{3}H$
($^{3}He$) with three quarks of an additional proton (neutron) again in
accordance with the above rules. Since each quark in each of the four
nucleons is coupled in a pair with a quark of an adjacent nucleon, current
quark configurations are totally suppressed and only constituent quark
configurations are realized in $^{4}He.$ Starting from $^{4}He$ all nuclei
possess 3D-crystal -- like structure. Indeed, planes of oscillations of
two protons and two neutrons are located on opposite faces of an
octahedron with common vertex. In this geometrical configuration four
nucleons are in an $s-$ state that corresponds to the first {\it s}--shell
of the shell model. Next, the $p-$ shell represents octahedron of bigger
size with two $^{3}He-$ triangles instead of protons and two $^{3}H-$
triangles instead of neutrons. The triangles are located parallel to empty
faces of the $^{4}He-$ octahedron, the free quark ends of these triangles
coupled as in the $^{4}He-$ octahedron. This octahedron with the nested
$^{4}He-$ octahedron represents the nucleus of $^{16}O.$ The next shell
with principal number $n=2$ is constructed in the same manner, extending
triangles beforehand by adding a row of three protons to the row of two
neutrons in $^{3}H$ and a row of three neutrons to the row of two protons
in $^{3}He.$ Again, these triangles are located in couples on opposite
faces of an octahedron parallel to unoccupied faces of the nested $p- $
octahedron. Construction of the next shells is performed in the same
manner by extending triangles with new rows of neutrons and protons. When
building the shells for $n>2$ one needs to take into account the
predominance of neutron number over proton number. It can be shown that at
fixed distances between nucleon centers of mass ($\sim 2$ $fermi)$ the
nucleons are arranged into a face--centered cubic lattice. It turns out
that at nucleonic degrees of freedom our quark model of nuclear structure
is identical to the lattice model formulated by N. Cook and V. Dallacasa
more than twenty years ago \cite {cook3} and called the FCC
(face--centered--cubic)--lattice model. They demonstrated that it brought
together shell, liquid-drop and cluster characteristics, as found in the
conventional models, within a single theoretical framework. Unique among
the lattice models, the FCC reproduces the entire sequence of allowed
nucleon states as found in the shell model. Manifestation of a crystalline
structure of nuclei has been observed in the diffraction pattern of
scattering of $\alpha $-particles on nuclei \cite{led}.

The above picture of quark rearrangements in multinucleon systems
resulting in considerable suppression of current quark configuration leads
to essential modifications of the nucleon properties inside nuclei. The
suppression of current quark configurations inside nuclei manifests in
various observable effects: the old ''EMC--effect'' \cite{emc}, color
transparency breaking in large angle quasielastic $pp- $ scattering off
nuclei, suppression of high transverse momenta, jets and $J/\psi $s in
heavy ion collisions at high energies. According to our model there are
holes (depression of hadronic matter density distribution) in the centers
of three-- and four--nucleon systems. This effect has been observed
experimentally \cite{hole}. And one more important corollary: all nuclei
are non--spherically symmetric, even nuclei with magic numbers. In
conclusion, the following consequences of the proposed approach should be
emphasized :

\begin{itemize}
\item  quarks and gluons inside hadrons and nuclei are strongly correlated;
\item  there are no strings stretching between quarks inside hadrons;
\item  strong interactions of quarks inside hadrons and nuclei are nonlocal;
\item  nuclei possess crystal--like structure.
\end{itemize}

I am very grateful to Norman Cook for help and discussions.

This research was partly supported by the Russian Foundation of Basic
Research, grant 01-07-90144.

\end{document}